\documentclass[a4paper,11pt]{article}
\usepackage{pos}

\title{Design and Expected Performance of the IceCube-Gen2 Surface Array and its Radio Component}
\ShortTitle{IceCube-Gen2 Surface Array}

\manuallySeparateAuthors 
\author*[a,b]{Frank~G.~Schr\"oder} 
\author[c]{ for the IceCube-Gen2 Collaboration}

\affiliation[a]{Bartol Research Institute, Department of Physics and Astronomy, University of Delaware, Sharp Lab, 104 The Green, Newark, DE 19716, USA}
\affiliation[b]{Institute for Astroparticle Physics (IAP), Karlsruhe Institute of Technology (KIT), Postfach 3640, 76021 Karlsruhe, Germany}
\affiliation[c]{Full author list at \url{https://icecube.wisc.edu/collaboration/authors/}}

\emailAdd{fgs@udel.edu}

\abstract{IceCube-Gen2, the next generation of the IceCube Neutrino Observatory at the South Pole, will consist of three co-located arrays: a deep Optical Array and a more shallow and larger Radio Array for neutrino detection in the ice, and a Surface Array above the footprint of the Optical Array. The Surface Array will be comprised of hybrid stations featuring elevated radio antennas and scintillation detectors, following the design of a prototype station successfully operating at the South Pole since 2020. Besides providing a veto for neutrino detection, the Surface Array will make IceCube-Gen2 a unique laboratory for cosmic-ray air showers. Compared to the current IceCube detector with its IceTop surface array, the aperture for coincident air-shower measurements detected by both, the deep optical and surface arrays, will increase by about a factor of 30. In addition to particle physics questions, such as the production of PeV muons and neutrinos in prompt decays, these surface-deep coincidences will be used to target astrophysical questions of the most energetic Galactic cosmic rays. The combination of particle and radio measurements at the surface and high-energy muons measured in the ice promises unprecedented accuracy for the mass composition in the energy range of the presumed Galactic-to-extragalactic transition – complementing the multimessenger science case of IceCube-Gen2.
This proceeding provides an overview of the IceCube-Gen2 Surface Array and, in particular, its radio component.}

\FullConference{%
  9th International Workshop on Acoustic and Radio EeV Neutrino Detection Activities - ARENA2022\\
  7-10 June 2022\\
  Santiago de Compostela, Spain}

\begin{document}
\maketitle

\section{Introduction}
IceCube-Gen2, the next generation extension of the IceCube Neutrino Observatory at the South Pole \cite{IceCube-Gen2:2020qha}, will feature three arrays: a deep Optical Array enlarging IceCube's current deep array by an order of magnitude in volume, a much larger Radio Array for ultra-high-energy neutrinos, and a Surface Array comprised of elevated scintillation and radio detectors above the footprint of the Optical Array. 
The Radio Array is described in a dedicated contribution at this conference \cite{Nelles:2023b3} and must not to be confused with the radio component of the Surface Array.
The Surface Array serves as a veto to improve the sensitivity to high-energy neutrinos of several $100\,$TeV to PeV, and -- beyond that -- enables a unique science case regarding the particle physics in cosmic-ray air showers and regarding the astrophysics of the most energetic Galactic cosmic rays. 
The science case is described in more detail in Ref.~\cite{Schroeder:2021t4}, and this proceeding will focus particularly on the aspects related to the elevated radio antennas of the Surface Array. 

\begin{figure}[b!]
\centering
\hspace{-1cm}
\includegraphics[width=0.685\linewidth]{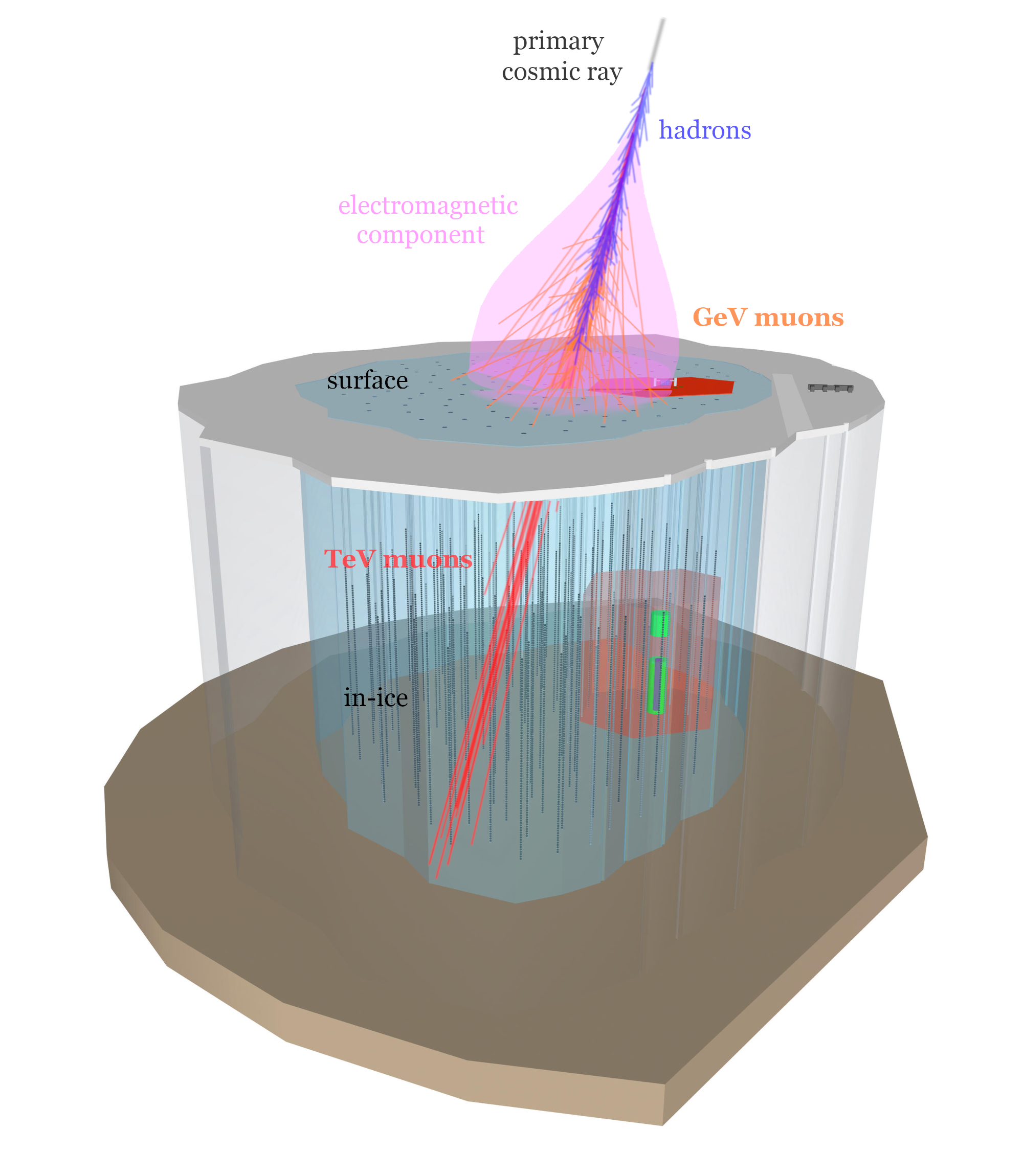}
\hspace{-1cm}
\includegraphics[width=0.41\linewidth]{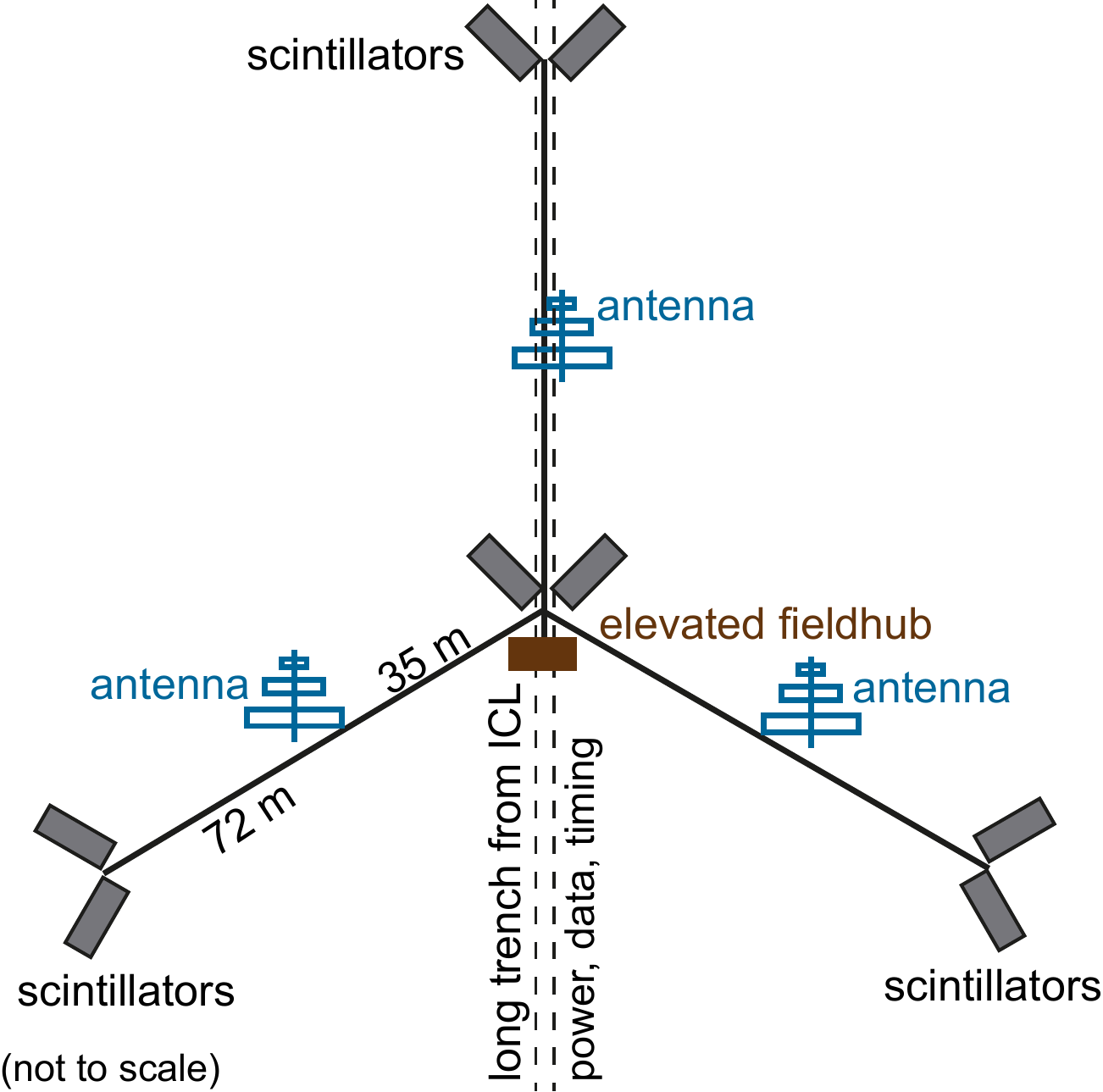}
\caption{Left: Sketch of an air shower detected by the IceCube (red) and IceCube-Gen2 (blue) surface and in-ice arrays. The electromagnetic component can be measured calorimetrically with the surface radio antennas. Additionally the electromagnetic particles and GeV mons will be observed by the particle-detector arrays at the surface. For those showers with favorable geometry, additionally the TeV muons around the shower axis will be detected by the optical detectors in the ice. Right: Sketch of a Gen2 surface station consisting of elevated scintillators and radio antennas connected to a fieldhub housing the station electronics.}
\label{fig:Gen2ArraySketch}
\end{figure}

The design of the IceCube-Gen2 Surface Array is inspired by a surface enhancement \cite{Haungs:2019ylq} planned to improve IceCube's existing surface array, IceTop \cite{IceCube:2012nn}. 
Each of the 120 strings of the IceCube-Gen2 Optical Array will feature one surface station consisting of eight scintillation and three radio detectors connected to a common DAQ in a central fieldhub that will also house electronics of the corresponding string (Fig.~\ref{fig:Gen2ArraySketch}).
By this design choice, the Surface Array profits from the communication, timing, and power infrastructure put in place for the Optical Array.
The scintillation panels will measure air-shower particles and provide a full-efficiency detection threshold of about $0.5\,$PeV for vertical proton showers. 
By elevating the detectors above the snow surface a flaw of IceTop is overcome, which continuously looses sensitivity due to increasing snow coverage. 
Moreover, the addition of radio antennas to the particle detector array will increase the accuracy of the air-shower measurements by providing a calorimetric energy \cite{Huege:2016veh,Schroder:2016hrv,PierreAuger:2016vya} and $X_\mathrm{max}$ \cite{LOPES:2012xou,Buitink:2014eqa,Apel:2014usa,Tunka-Rex:2015zsa} measurement in the energy range of the Galactic-to-extragalactic transition.

\section{Prototype Station at the South Pole}
Following earlier tests of prototype detectors, a full prototype station of the surface enhancement is in operation at the South Pole since 2020. 
It consists of eight scintillation panels with digitizers implemented directly in each panel, and three SKALA v2 \cite{SKALAv2} radio antennas, which are connected to the same central data-acquisition `TAXI' as the scintillators \cite{Haungs:2019ylq}. 
TAXI digitizes the radio signals upon a trigger, which is either a coincidence trigger from the scintillators for the measurement of air-shower induced radio signals or a software-generated trigger for periodic measurement of the radio background at the South Pole.

Compared to several other radio arrays for air showers, the chosen antenna type enables a relatively high frequency band of $50-350\,$MHz.
This band provides a better signal-to-noise ratio, than the traditional band of $30-80\,$MHz, at least for those showers that are detected inside-of the Cherenkov ring where the radio emission extends up to several $100\,$MHz \cite{BalagopalV:2017aan}.
The choice of the SKALA antennas thus enables a relatively low detection threshold, which currently is around $30\,$PeV for the prototype station \cite{IceCube:2021epf}.
This threshold will be even further reduced by improvements of the DAQ electronics reducing the internal noise and by the application of neural networks for the detection of air-shower pulses \cite{Rehman:2021nw}.
Overall, the prototype station has thus successfully demonstrated that the SKALA v2 antennas are a suitable choice to study cosmic rays in the transition region from Galactic-to-extragalactic cosmic rays.

\begin{figure}[t]
\centering
\includegraphics[height=5.8cm]{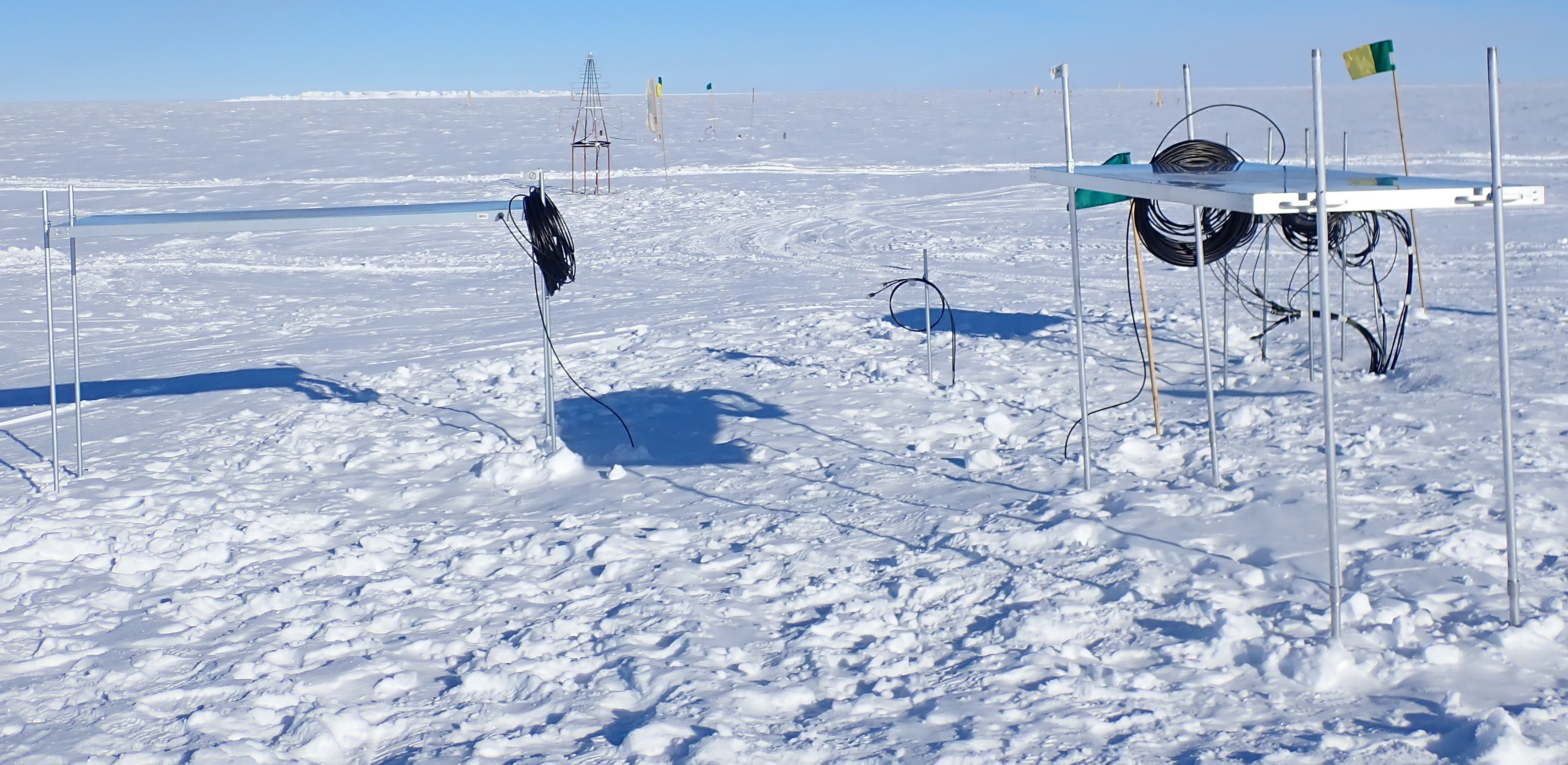}
\hfill
\includegraphics[height=5.8cm]{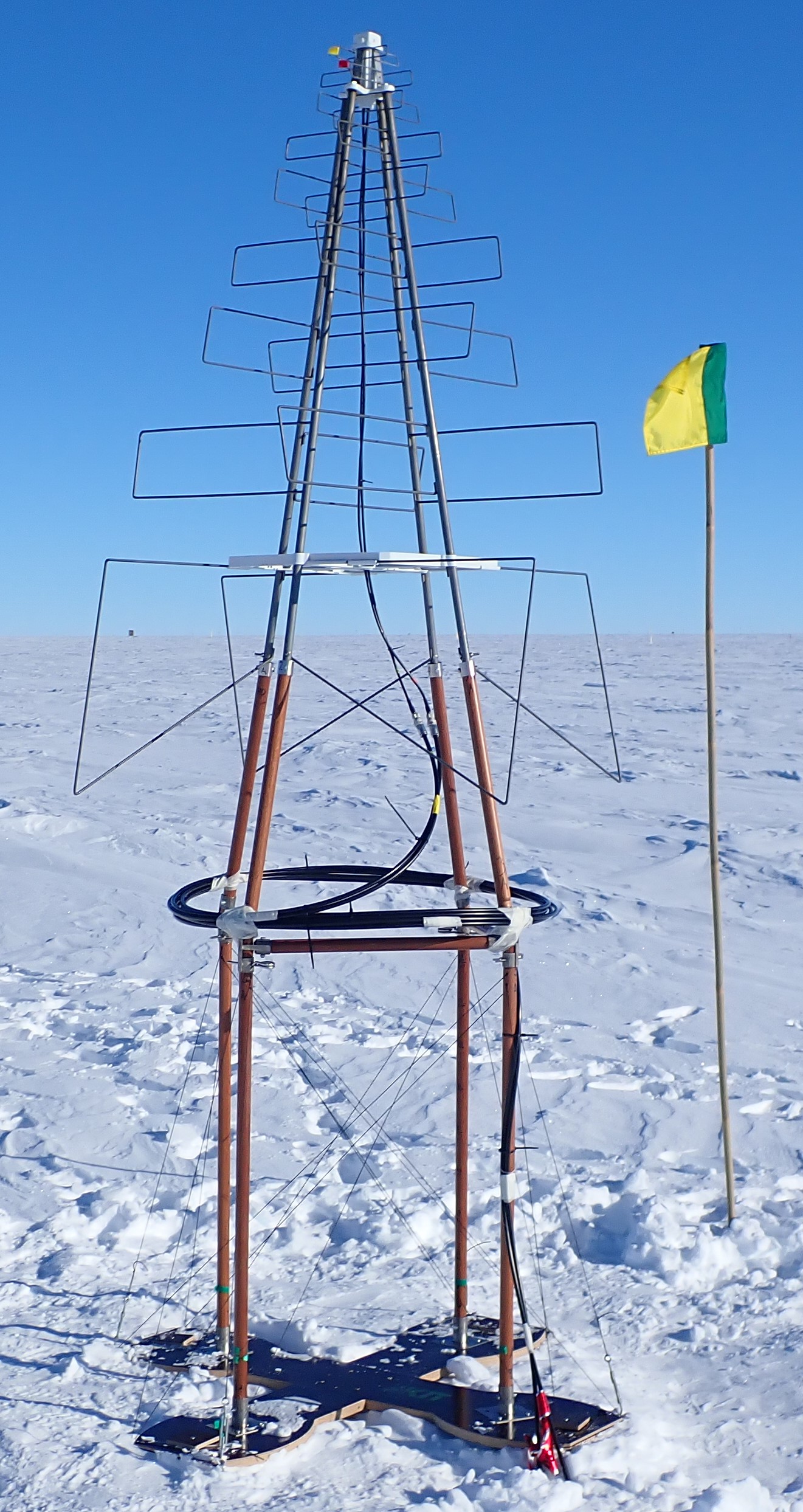}
\caption{Left: two scintillators and one antenna in the background of the prototype station at the South Pole deployed in January 2022. Right: close-up photo of the antenna.}
\label{fig:PrototypeStationPhotos}
\end{figure}

Moreover, the prototype station has demonstrated that the detector design is well fit to the conditions at the South Pole.
Both, the scintillation detectors and the radio antennas, are deployed on mounts that elevate the detectors by more than $1\,$m above the surface (Fig.~\ref{fig:PrototypeStationPhotos}). 
This design avoids snow coverage of the detectors. 
Three years of operation at the South Pole show that no additional snow accumulation is caused by the detectors beyond the unavoidable general increase of the snow height: the observed rise of the snow level under the detectors and around them corresponds to the approximately $20\,$cm per year that is observed everywhere around the South Pole. 
Therefore, the scintillators and antennas have to be raised once every about 5 years, which is facilitated by the extendable design of the mounts.
Finally, the detectors have been operating successfully under all weather conditions throughout the years, and are thus chosen as baseline design for the IceCube-Gen2 Surface Array.

\section{Simulation Study for the Radio Component of the Surface Array}
The IceCube-Gen2 Surface Array is thus planned to be comprised of about 160 such stations: one for each of the 120 optical strings of IceCube-Gen2, the other stations mostly for the surface enhancement of the current IceTop array and a few more to provide a smooth transition between these two parts of the surface array. 
Over the largest part of the array, the spacing between the stations will hence be the same as the approximately $240\,$m spacing between the optical strings, i.e., the surface antenna density will be about $50-60$ antennas per km$^2$.

The detection efficiency of the radio component of the IceCube-Gen2 Surface Array has been assessed using CoREAS simulations of proton and iron primaries over a wide range of energies and zenith angles \cite{Abbasi:2021oc}. 
These simulations calculate the radio signal on ground on a star-shape pattern, so the core position relative to the array can be varied and the signal at each antenna position interpolated from the star-shape pattern \cite{IceCube:2022dcd}. 
Motivated by measurements of the prototype station which show that, apart from some human-made background from the South Pole infrastructure, the underlying noise level is consistent with the expected Galactic noise, a background consisting of Galactic noise and the thermal noise of the low-noise amplifier has been added to the CoREAS signal.
Then the responses of the antennas, the analog signal chain, and the TAXI DAQ are applied to the simulated radio pulses to mimic signals as they would be detected by the IceCube-Gen2 Surface Array.

\begin{figure}[t]
\centering
\vspace{-1mm}
\includegraphics[width=0.999\linewidth]{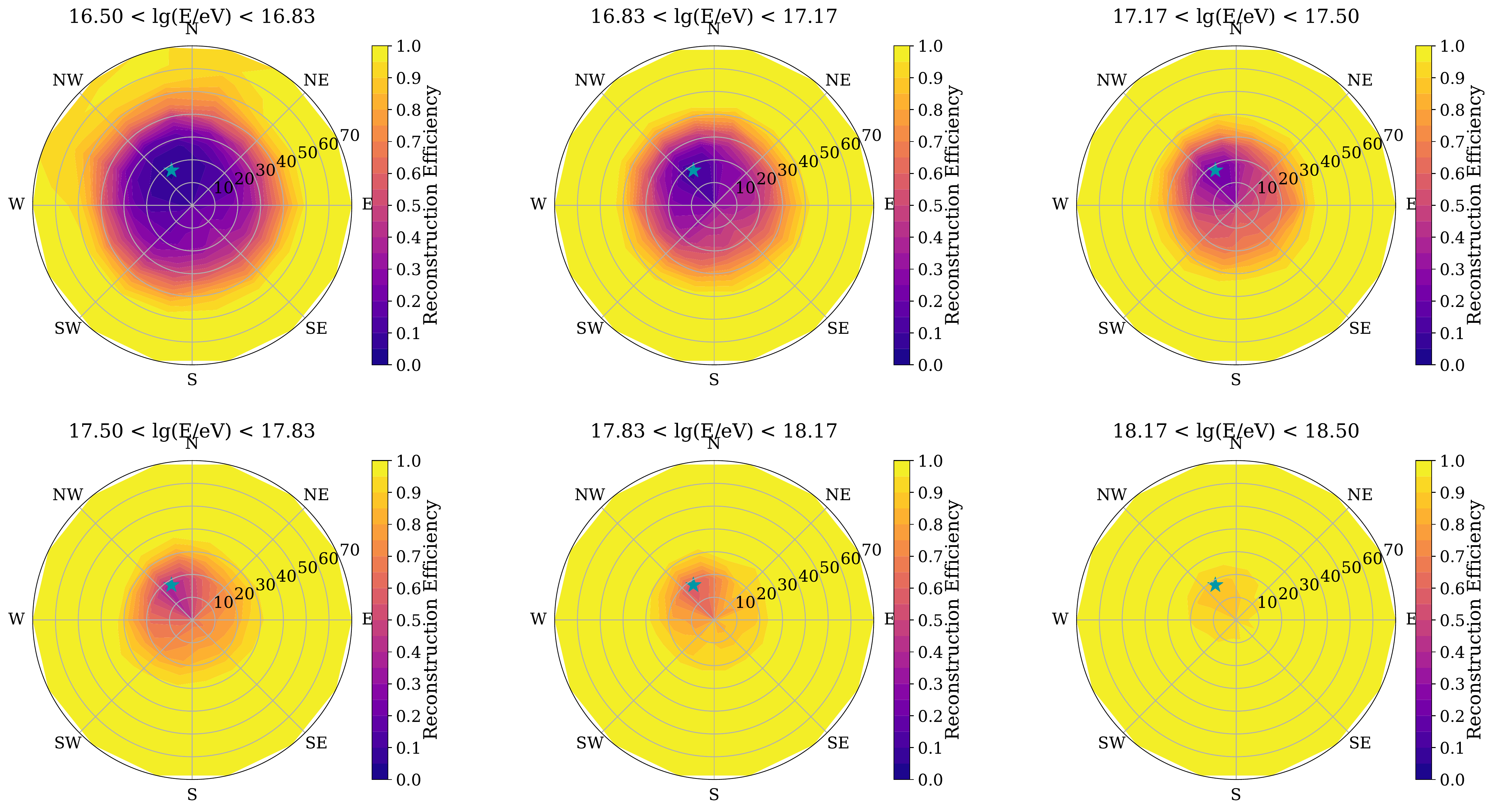}
\caption{Efficiency of the radio component of the IceCube-Gen2 Surface Array derived from CoREAS simulations by assuming a simple detection threshold for the radio signal in individual antennas. Full efficiency is expected for inclined showers over a wide range of azimuth angles starting from a few $10\,$PeV and extending to almost the full sky in the EeV energy range. Even for arrival directions parallel to the geomagnetic field (indicated as a star), a fraction of the showers will be detected.}
\label{fig:ThresholdOverEnergy}
\end{figure}

For these signals including the complete detector response, a detection threshold is defined by a signal-to-noise ratio that removes $99\,\%$ of the background events in individual antennas. 
Such a simple threshold is sufficient because the antennas will not be self-triggered, but instead be triggered by the scintillators of the Surface Array as well as by the Optical Array in the ice, when these detect an air-shower or a muon track (which in most cases is the product of an air shower).
Given the size of the array with more than 100 antennas, it is still likely that some false-positive antennas are present in any externally triggered event, but in most cases they can easily be removed because they will represent outliers in the lateral and time distributions of the radio signal of the air showers. 
Furthermore, more sophisticated detection methods, such as neural networks \cite{Rehman:2021nw}, will reduce the false-positive rate compared to the simple threshold criterion used here.
To avoid a significant contamination of this simulation study by antennas with upward fluctuations of noise, we require at least three antennas over threshold. 
In addition, the reconstructed arrival direction from a plane fit through the antenna signals must match the true direction within five degrees (where the reconstruction of the arrival direction takes into account the signal strengths and ignores distant antennas to reduce the impact of false-positive signals).
Later when the real array is deployed, a similar cross-check can be applied against the scintillator reconstruction of the same air showers. 
Applying these detection criteria, we can thus determine the detection efficiency including the efficiency of this simple direction reconstruction as function of energy and zenith angle (Fig.~\ref{fig:ThresholdOverEnergy}).

As can be seen from the yellow areas in the figure, full efficiency is expected for arrival directions from a significant fraction of the sky starting at a few $10\,$PeV. 
Showers between $50^\circ$ and $70^\circ$ zenith angle will feature the lowest threshold, where this sky region happens to include the Galactic Center which is constantly visible from the South Pole at $61^\circ$ zenith angle. 
In contrast to the current IceCube detector with its planned surface enhancement, due to the larger extension of the Surface Array, coincidences with the in-ice Optical Array will occur also for some of these inclined showers. 
Therefore, even among these inclined showers we expect coincidences with the deep Optical Array that will help for practically all science goals, including gamma-hadron separation for the search of photons from the Galactic Center, which is one of the exciting side goals, as it provides a chance to directly discover the source of the most energetic Galactic cosmic rays.

\begin{figure}[t]
\centering
\vspace{-3mm}
\includegraphics[width=0.6\linewidth]{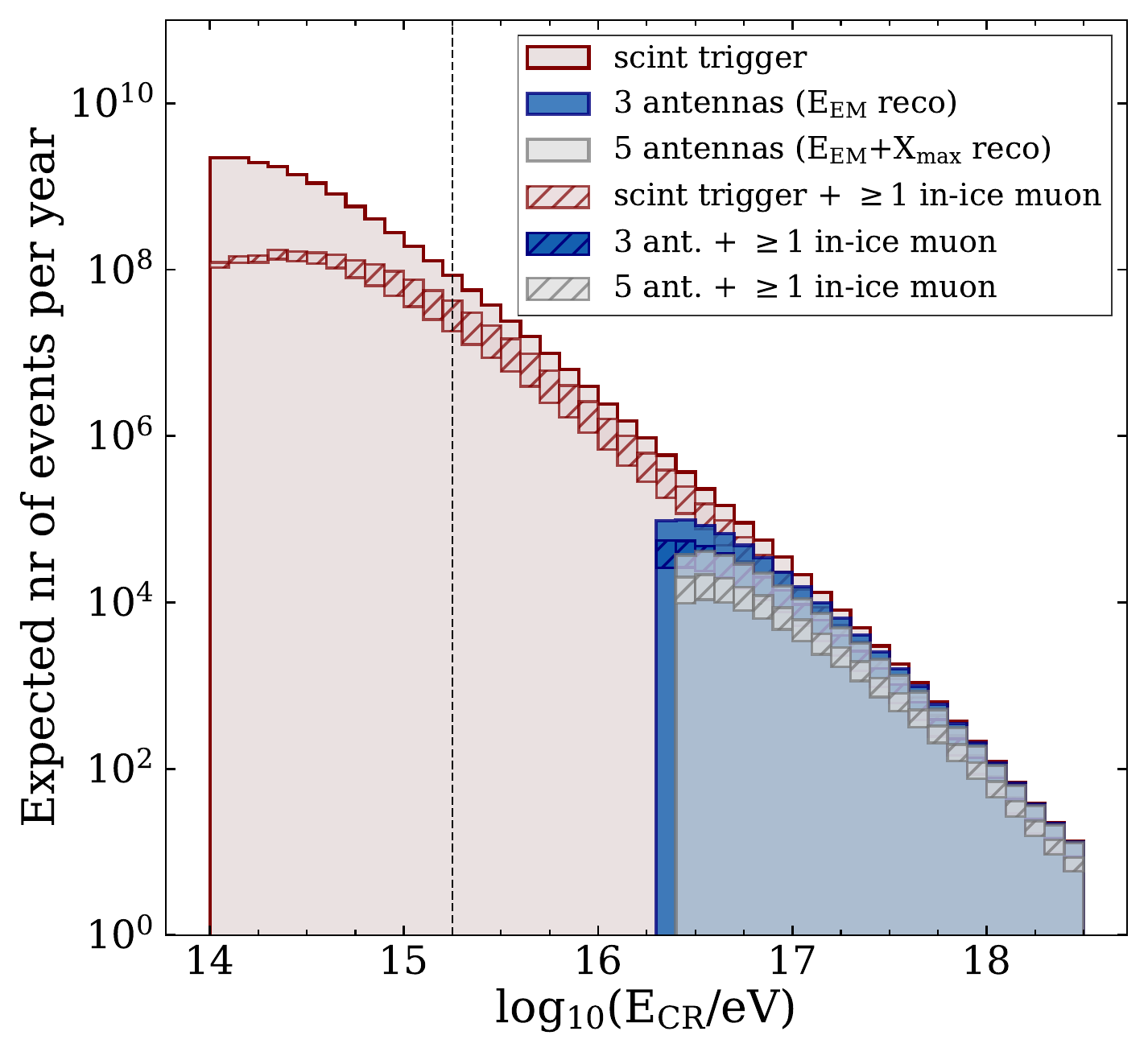}
\vspace{-1.3mm}
\caption{Expected yearly statistics of cosmic-ray air-shower events triggered by the scintillators simulated using a 50/50 mix of iron and proton primaries, and the portion of those events which contain at least one high-energy muon reaching the in-ice Optical Array and/or a detectable radio signal. Above a few $10\,$PeV, a sizeable fraction of the cosmic-ray events will have both, radio signals and at least one high-energy muon, in addition to the measurement by the scintillator array. Consequently, IceCube-Gen2 will detect a significant number of such multi-hybrid events in the energy range of the presumed Galactic-to-extragalactic transition.}
\label{fig:eventStatistics}
\vspace{-0.3mm}
\end{figure}

Generally, air showers measured in coincidence between the radio component of the Surface Array and the deep Optical Array will be most valuable for the physics goals. 
Taking the increased area and the increased zenith range of IceCube-Gen2 together, the aperture for coincidences between the Surface and Optical Arrays will be about 30 times larger in IceCube-Gen2 compared to current IceCube.
The simultaneous measurement of TeV muons in the ice and the radio measurement of the electromagnetic component will enable new insights into the particle physics of these air showers. 
Furthermore, having a precise calorimetric measurement of the size of the electromagnetic component, $X_\mathrm{max}$, and the TeV muons will enable a higher accuracy for the evolution of the mass composition over the cosmic-ray energy than achieved by current experiments in this energy range. 
This is because at a given energy, which can be accurately reconstructed from the radio signal \cite{PierreAuger:2016vya,Bezyazeekov:2018yjw,LOPES:2021ipp}, $X_\mathrm{max}$ and muons feature complementary mass sensitivity.
Figure~\ref{fig:eventStatistics} shows the expected statistics of certain classes of events, including these most valuable multi-hybrid events, which will make IceCube-Gen2 a world-unique detector for the most energetic Galactic cosmic rays.

\section{Conclusion}
By the combination of a Surface Array including radio antennas and an in-ice Optical Array, IceCube-Gen2 will constitute a unique laboratory to study the particle and astrophysics of cosmic rays in the energy range from below a PeV to several EeV. 
In particular, for energies above several $10\,$PeV, i.e., in the energy range where the most energetic Galactic cosmic rays and the transition to an extragalactic origin are presumed, the combination of energy and $X_\mathrm{max}$ measurements with the in-ice muon signal will enable a higher accuracy for the mass and new insights into the shower physics. 
Although such surface and in-ice coincidences occur already for the current IceCube detector, the aperture for these will increase by a factor of 30 with IceCube-Gen2 and include more inclined arrival directions for which the radio antennas have a lower full-efficiency threshold.

The science potential of the Surface Array includes its use as veto for neutrino measurements, the cross-calibration of the energy scale with other air-shower arrays \cite{Tunka-Rex:2016nto,Mulrey:2021e1}, and cosmic-ray physics.
In particular, IceCube-Gen2 will contribute to open questions regarding the muon puzzle in air showers, the study of prompt leptons, and the astrophysical questions related to the Galactic-to-extragalactic transition. 
By providing extended measurements of the cosmic-ray anisotropy together with more accurate measurements of the mass composition, IceCube-Gen2 will study the origin of the most energetic Galactic cosmic rays and constrain scenarios for the transition to cosmic rays of extragalactic origin.
The importance of combining the IceCube-Gen2 Optical Array with a Surface Array has consequently been outlined also in the recent Astro2020 \cite{SchroderAstro2020} and Snowmass \cite{Coleman:2022abf} surveys. 
The radio antennas of the IceCube-Gen2 Surface Array are hence critical to provide the additional accuracy needed to achieve these important science goals in cosmic-ray physics -- complementing the multi-messenger science of the IceCube-Gen2 Neutrino Observatory.

\bibliographystyle{JHEP}
\bibliography{references}
\vspace{5mm}
\noindent
\footnotesize{The preparations for IceCube-Gen2 are supported by several funding agencies. Specific acknowledgement for this proceeding: This project has received funding from the European Research Council (ERC) under the European Union's Horizon 2020 research and innovation programme (grant agreement No 802729). Parts of the presented work have been supported by the U.S. National Science Foundation-EPSCoR (RII Track-2 FEC, award ID 2019597).}

\end{document}